\documentclass{aa}
\usepackage{txfonts}
\usepackage{epsf}

\begin{document}

\title{GaaP: PSF- and aperture-matched photometry using shapelets}
\author{Konrad Kuijken}
\institute{Leiden Observatory, PO Box 9513, 2300RA
  Leiden, The Netherlands}
\date{Received <date> / Accepted <date>}

\abstract{}{We describe a new technique for measuring accurate galaxy
  colours from images taken under different seeing conditions.  }{The
  method involves two ingredients. First we define the
  Gaussian-aperture-and-PSF flux, which is the Gaussian-weighted flux
  a galaxy would have if it were observed with a Gaussian PSF. This
  theoretical aperture flux 
  is independent of the PSF or pixel scale that the galaxy was
  observed with. Second
  we develop a procedure to measure such a `GaaP' flux from observed,
  pixellated images. This involves modelling source and PSF as a
  superposition of
  orthogonal shapelets. A correction scheme is also described, which
  approximately corrects for any residuals to the
  shapelet expansions.}{A series of tests on simulated images shows
  that with this method it is possible to reduce systematic errors in
  the matched-aperture fluxes to a percent, which makes it useful for
  deriving photometric redshifts from large imaging surveys.}{}

\maketitle

\section{Introduction}

Accurate measurement of colours of astronomical sources is important
for many applications. In this paper we describe a new technique for
colour measurement that is specifically geared towards poorly-resolved
galaxy images in wide-area optical surveys. It is often the case that
images taken through different filters have different image quality
(different seeing or pixel scale, for example), and it is important to
correct the photometry for these effects. An important application is
the measurement of photometric redshifts of galaxies (e.g.,
Baum~\cite{Baum1957}) from multi-band imaging surveys such as SDSS
(Adelman-McCarthy et al.~\cite{sdssdr4}), FIRES (Labb\'e et
al.~\cite{fires1}; F\"orster Schreiber et al.~\cite{fires2}), FDF
(Heidt et al.~\cite{fdf}), CFHTLS (Brodwin et al.~\cite{cfhtls3}), or
the up-coming KIDS survey on the VST (Kuijken et al., in preparation).

Ideally, the total flux for each source can be measured in each band,
after which the colours follow simply. But in practice we are limited
to measuring aperture fluxes, because of pixel noise and confusion by
neighbouring sources. Finite aperture fluxes are affected by smearing
by the point spread function (PSF) and by pixellation.

To measure accurate colours does not require good measurements of the
total flux of a source: it is sufficient to determine the flux of a
particular intrinsic part of the source in different wavelength bands,
and compare those. The aperture need not have sharp edges: any weight
function will do.  The technical problem to solve is then how to make
sure that the same intrinsic part of the source is measured in each
band, when all that is available is a set of convolved images of the
source, each with a different PSF.

An often-used solution is to smooth all images to the same effective
PSF. This involves convolution of the sharpest images in a multi-band
set with appropriate convolution kernels, resulting in smoothed images
that have the same PSF as the worst-seeing image of the
set. Matched-aperture photometry on these images then yields proper
colours. A practical disadvantage of this method is that all images
need to be brought together in order to compute the proper convolution
kernels---in case of a large multi-band survey this is not a simple
task. It is also computationally expensive to convolve all pixels of
an image with a spatially variable kernel, particularly if most of
these pixels will in the end not be part of any source. 

This paper describes a different approach which is considerably more
efficient and no less accurate.  Rather than convolving images to each
other's PSF, we encode the detected sources in each image as a sum of
orthogonal basis functions (`shapelets', Refregier
\cite{Refregier2003}, henceforth R03). The same is done for the
PSF. From this point on we only manipulate the sources, removing the
need to convolve all pixels.  For each source we then construct a 'PSF
cleaning' operator with which to compute the aperture flux that would
be measured with a standard (Gaussian) PSF. By standardizing on a
'clean' PSF we avoid the need to know the 'dirty' PSFs of all bands at
once, which means that images can be analysed quite independently.

This paper is organized as follows.  Section~\ref{sec:def} defines the
aperture flux we will use. Section~\ref{sec:shapelets} summarizes the
shapelets formalism, and we show there how to use it to compute our
aperture fluxes for shapelet objects. An error analysis is also
given. In section~\ref{sec:tests} we present the results of many
simulations, and derive several technical improvements which take into
account the residuals to the shapelet description for the observed
sources and PSFs. Tests of the noise propagation and a discussion of
the choice of aperture radius are treated in
section~\ref{sec:noise}. After a short discussion
(Section~\ref{sec:discussion}) we summarize the paper in
Section~\ref{sec:summary}.

\section{An aperture flux measure independent of PSF}
\label{sec:def}

We define the 'Gaussian-aperture-and-PSF' (GaaP) flux $F_q$ of
aperture radius $q$ as the Gaussian-weighted flux the source would
have if it were observed with a Gaussian PSF of the same width as the
weight function:
\begin{equation}
F_q\equiv \int d{\bf r}\, e^{-r^2/2q^2} 
\int d{\bf r'} S({\bf r'})
{e^{-({\bf r}-{\bf r'})^2/2q^2}\over2\pi q^2}
\label{eq:fq1}
\end{equation}
where $S$ is the intrinsic (pre-PSF, pre-pixelization) light
distribution of the source.  We will show below how we can estimate
$F_q$ from observed (PSF-convolved, pixelated) images of the source.

Changing the order of integration in eq.~\ref{eq:fq1} and some simple
algebra shows that $F_q$ is equivalent to the integral
\begin{equation}
F_q=\int d{\bf r}\, \frac12 e^{-r^2/4q^2} S({\bf r}),
\label{eq:fq2}
\end{equation}
i.e., half of the weighted flux of the intrinsic source $S$ with a Gaussian
weight function of dispersion $2^{1/2}q$.  

\begin{figure}
\epsfxsize=0.49\hsize\epsfbox{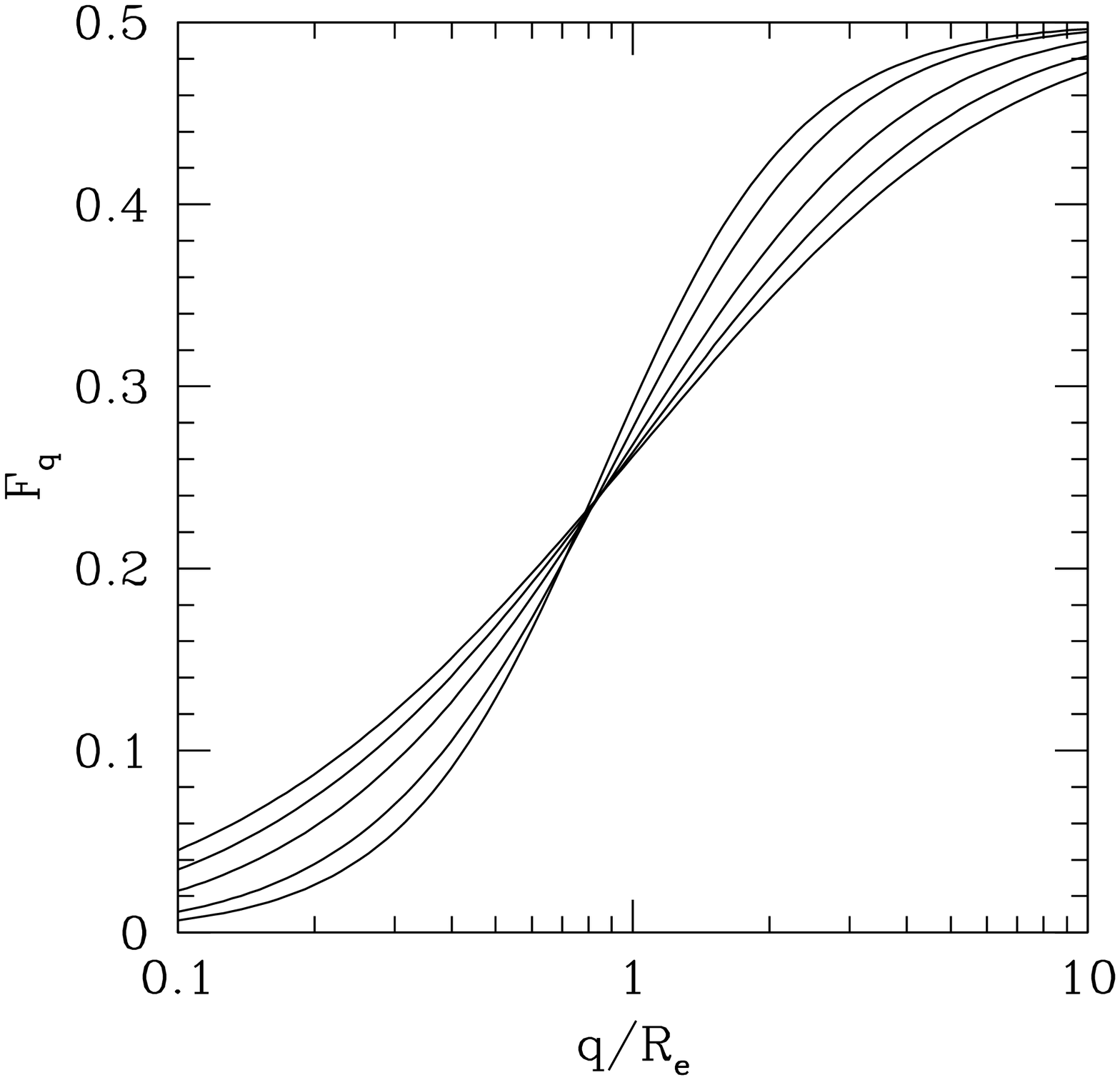}
\epsfxsize=0.49\hsize\epsfbox{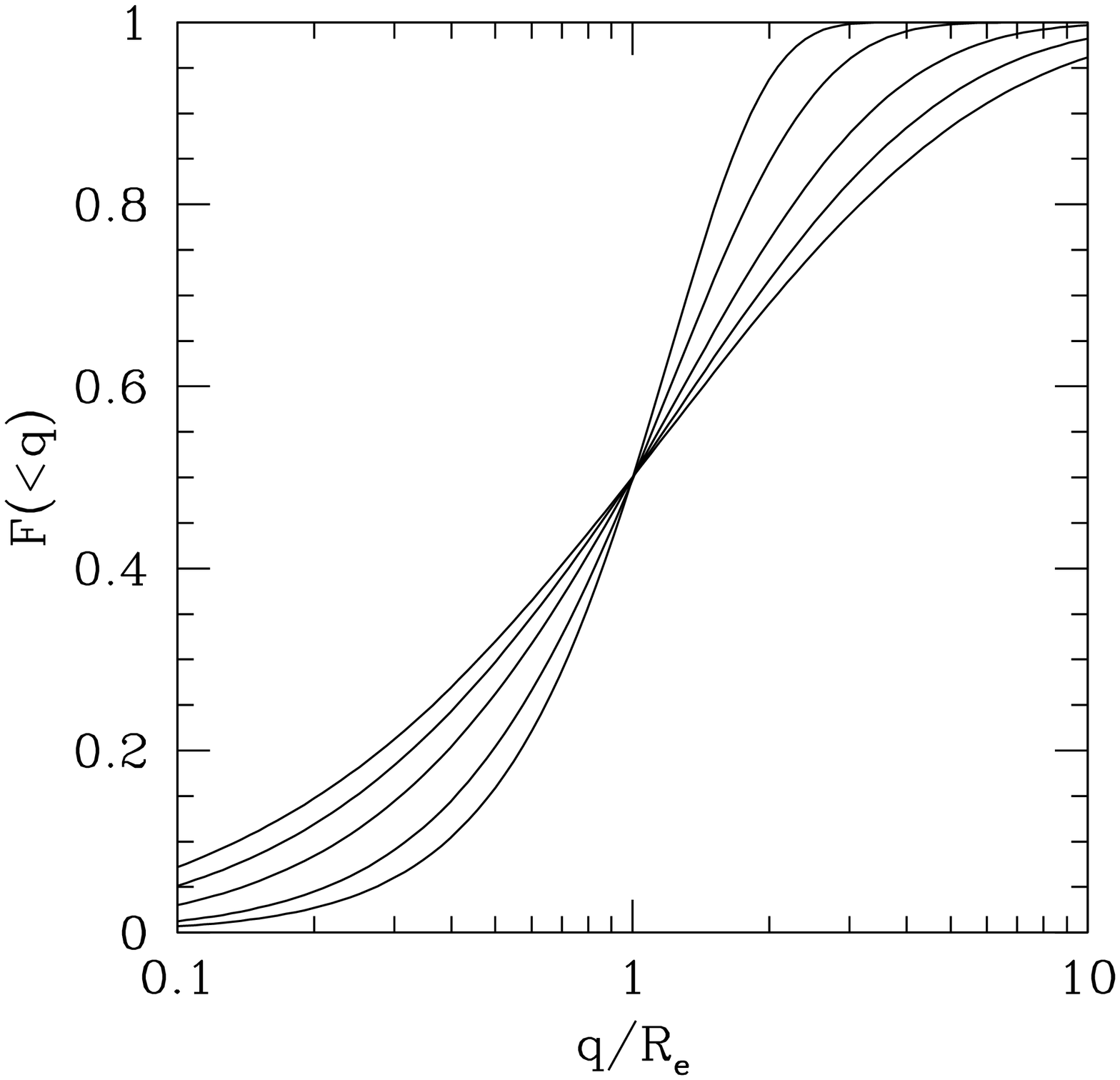}
\caption{Left: The GaaP aperture flux for different Sersic profiles of
  effective radius 1, and total flux 1, as a function of aperture
  radius $q$. From steepest to flattest, Sersic index 0.5 (Gaussian),
  1 (exponential), 2, 3 and 4 (de Vaucouleur) are plotted. Right: the
  top-hat aperture fluxes within radius $q$ for the same profiles.}
\label{fig:gaap-sersic}
\end{figure}

The GaaP flux for Sersic model light profiles is plotted in
Fig.~\ref{fig:gaap-sersic}, and compared with the top-hat aperture
flux. As expected (because the GaaP flux includes some smoothing), the
GaaP flux has a flatter curve of growth than the straight aperture
flux, and is less sensitive to differences in galaxy type.

\section{Shapelets}
\label{sec:shapelets}

In the shapelets formalism (R03) images are expressed as a truncated
sum of orthonormal Gauss-Hermite terms. In this paper we follow the
notation used in Kuijken (\cite{Kuijken2006}, henceforth K06), where
an application of shapelets to weak gravitational lensing is
described.

The shapelet basis function of scale
$\beta$ and $x,y$ orders $a,b$ is
\begin{equation}
B^{ab}(x,y)=k^{ab}\beta^{-1}
 e^{-[(x-x_c)^2+(y-y_c)^2]/2\beta^2} H^a(x/\beta) H^b(y/\beta),
\end{equation}
where $x_c$ and $y_c$ are
 the center of the expansion, $B^{ab}$ is the basis function of order
 $(a,b)$, $H^a$ is the Hermite polynomial of order $a$ and 
\begin{equation}
k^{ab}=1/\sqrt{\pi a!b!2^{a+b}}
\end{equation}
 is
 a normalization constant chosen so that 
 $\left(B^{ab}\right)^2$ integrates to one. 

 An image of a source, or a point spread function, can be expressed as
 a sum of such basis functions with coefficients $s_{ab}$. This can be
 done in two, not quite equivalent ways: by integration or by fitting.
 Integration is more straightforward in principle. The $B^{ab}$ are an
 orthonormal basis, so the coefficient of each term is simply the
 integral of the image times this basis function. Since the image is
 pixellated, however, performing this integral numerically is
 fundamentally inaccurate since it needs to be approximated by a
 finite sum. Fitting has the advantage that it is well-defined: one
 minimizes $\chi^2$ over all coefficients, which yields a description
 of an image that fits the fluxes in each image pixel. Formally, the
 best-fit coefficients therefore describe the pixel-convolved image,
 which means that when image and PSF are both fitted in the same way,
 their deconvolution should be independent of the pixel
 scale.\footnote{A third possibility is to fit the image as a
   superposition of pixel-convolved shapelet basis functions. This
   yields a description of the pre-pixellation image. A disadvantage
   of this approach is that the basis functions are no longer
   orthogonal, implying that the coefficients depend on the order at
   which the basis function set is truncated.}  In this paper we
 determine shapelet coefficients by least-squares fitting to the pixel
 values.

Once the shapelet coefficients have been determined they can be
operated on to calculate the effects of convolution, translation,
magnification, shear, etc, as well as a range of photometric
quantities. For the present application the relevant operations are
convolution, and weighted-aperture photometry.

Convolution of a source by a PSF is a linear operator and can be
represented by the action of a PSF matrix ${\bf P}$. The elements of
${\bf P}$ depend on the shapelet coefficients $p_{ab}$ and scale
$\beta_\mathrm{psf}$ of the PSF, on the shapelet scale
$\beta_\mathrm{in}$ of the source being convolved, and on the shapelet
scale $\beta_\mathrm{out}$ of the result.  For a model source whose
shapelet coefficients are $m_{ab}$, the convolution with the PSF can
be written as
\begin{eqnarray}
\nonumber
&\left(\sum_{a,b} p_{ab}B^{ab}_{\beta_\mathrm{psf}}\right)
\otimes
\left(\sum_{a,b} m_{ab}B^{ab}_{\beta_\mathrm{in}}\right)
\qquad\qquad\qquad
\\
&=
\sum_{a_1,b_1}\left(
\sum_{a_2,b_2} 
{\bf P}_{a_1b_1a_2b_2}(\beta_\mathrm{in},\beta_\mathrm{out})m_{a_2b_2}
\right)
B^{a_1b_1}_{\beta_\mathrm{out}} .
\end{eqnarray}
 The coefficients of the PSF matrix are
\begin{equation}
{\bf P}_{a_1b_1a_2b_2}(\beta_{\rm in},\beta_{\rm out})=
\sum_{a_3,b_3}
C_{a_1a_3a_2}^{\beta_{\rm out}\beta_{\rm psf}\beta_{\rm in}}
C_{b_1b_3b_2}^{\beta_{\rm out}\beta_{\rm psf}\beta_{\rm in}} 
p_{a_3b_3}
\label{eq:cmnl}
\end{equation}
where the elemental convolution matrix
$C_{mnl}^{\beta_3\beta_1\beta_2}$ expresses the convolution of two
one-dimensional shapelets of scales $\beta_1$ and $\beta_2$ as a new
shapelet series with scale $\beta_3$. A recurrence relation for
$C_{mnl}$ is given in R03.

If the image of the source $S$ is expressed as a shapelet series, with
scale radius $\beta$, then the GaaP flux defined in \S\ref{sec:def}
can be calculated analytically for each $S=B^{ab}$: only those terms
for which the orders $a$ and $b$ are both even contribute, and in this
case the $x$ and $y$ integrals in eq.~\ref{eq:fq2} separate as
$F_q=F_q^a F_q^b$ with
\begin{equation}
F_q^a=\int dx\,\frac12 e^{-x^2/4q^2} {H^a(x'/\beta)e^{-x'^2/2\beta^2}\over
\sqrt{\sqrt\pi 2^a a! \beta}}
\end{equation}
which, with the help of the result that, for even $n$,
\begin{equation}
\int_{-\infty}^\infty e^{-t^2}H^n(tx)dt=\sqrt\pi{n!\over(n/2)!}(x^2-1)^{n/2}
\label{eq:gr}
\end{equation}
(Abramowicz and Stegun~\cite{a+s}, eq.~22.13.17) reduces to
\begin{equation}
F_q^a={\pi^{1/4}\beta^{1/2}\over\sqrt{1+\beta^2/2q^2}}
\sqrt{(a-1)!!\over a!!}\left(2q^2-\beta^2\over2q^2+\beta^2\right)^{a/2}.
\label{eq:fqshapelets}
\end{equation}
(where $0!!=1!!=1$ and $n!!=n (n-2)!!$).
Hence the Gaussian-aperture, Gaussian-PSF flux of an object with
shapelet coefficients $s_{ab}$ at scale $q$ is simply the sum 
\begin{equation}
F_q(s)=\sum_{a,b\,\,\mathrm{even}}s_{ab}F_q^aF_q^b.
\label{eq:gaap}
\end{equation}

To calculate the GaaP flux of an object $S$ observed with
PSF $P$, we therefore 
\begin{enumerate}
\item
express both as shapelet series $s_{ab}$ and
$p_{ab}$, using scales $\beta$ and $\beta_\mathrm{psf}$ respectively;
\item
construct the PSF matrix ${\bf P}$ using eq.~\ref{eq:cmnl}, using
the same value $\beta$ as the input and output shapelet scale;
\item
calculate the shapelet expansion for the deconvolved image
${\bf P}^{-1}\cdot{\bf s}$;
\item
calculate the GaaP flux using eqs.~\ref{eq:fqshapelets} and
\ref{eq:gaap}. 
\end{enumerate}
The resulting expression is
\begin{equation}
F_q=\sum_{a_1,b_1\,\,\mathrm{even}}\ \ \sum_{a_2,b_2}\ \ 
{\bf P}_{a_1b_1a_2b_2}^{-1} F_q^{a_1}F_q^{b_1} s_{a_2b_2} .
\label{eq:straightfq}
\end{equation}
From this expression it is also straightforward to calculate the
variance of $F_q$: for background-limited imaging the standard errors
on the $s_{ab}$ are uncorrelated and, because of the normalization of
the basis functions, equal to the standard error $\sigma$ on the flux
in a pixel. Hence the variance of the GaaP flux is
\begin{equation}
\hbox{Var}\left(F_q\right)=
\sum_{a_2,b_2}
\left(\sum_{a_1,b_1\,\,\mathrm{even}}
{\bf P}_{a_1b_1a_2b_2}^{-1} F_q^{a_1}F_q^{b_1}\right)^2 
\sigma^2.
\label{eq:varfq}
\end{equation}


\begin{figure}
\epsfxsize=\hsize\epsfbox[20 150 575 575]{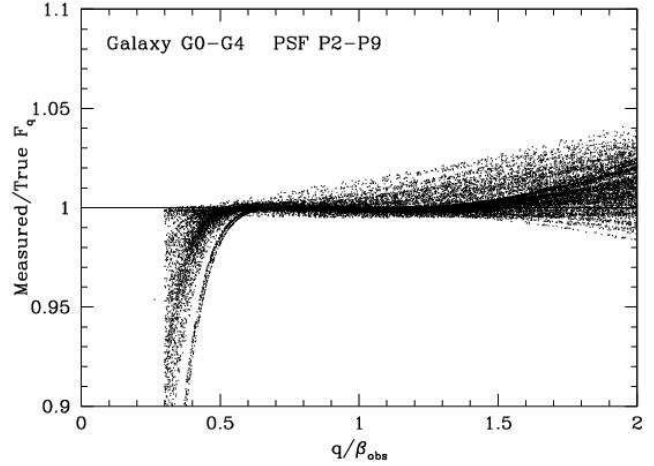}
\caption{The ratio of the measured GaaP flux to the true value, for a
  series of simulated Sersic profile model galaxies convolved with
  Moffat PSF profiles. Models are identified in Table~\ref{tab:types}. 
  The horizontal axis is the aperture radius $q$,
  divided by the shapelet scale used to describe the observed
  image. See the text for details. Shapelet order 8 was used. (Colour
  versions of this and similar figures indicating the parameter
  dependence of the results are included in the on-line materials.)}
\label{fig:firstrun}
\end{figure}

\begin{table}
\centering
\caption{PSF and Galaxy profiles used in the tests}
\label{tab:types}
\begin{tabular}{cl}
\hline\hline
PSF type & \\
\hline
0 & Pseudo-Airy (eq.~\ref{eq:pseudoairy})\\
1 & Pseudo-Airy with 10\% spikes\\
2 & Moffat, $m=2$\\
3 & Moffat, $m=3$\\
9 & Moffat, $m=9$ (nearly Gaussian) \\
&\\
Galaxy type & \\
\hline
0 & Gaussian\\
1 & Sersic, $n=1$ (exponential)\\
2 & Sersic, $n=2$\\
3 & Sersic, $n=3$\\
4 & Sersic, $n=4$ (de Vaucouleur)\\
5 & Spiral (eq.~\ref{eq:spiral})\\
\hline
\end{tabular}
\end{table}

\section{Tests and improvements}
\label{sec:tests}

To test this procedure, we performed the following simulations. First
we generated a series of model galaxy images with Sersic (\cite{Sersic1968})
profiles $\exp(-kr^{1/n})$. These were then convolved with Moffat profile
$(1+r^2/a^2)^{-m}$ PSFs, and
pixellated to yield simulated images. A corresponding pixellated PSF
image was generated in the same way. All images were built by
monte-carlo selection, as described in K06.

Each pixelated image and its PSF was then fitted with a shapelet
series to order 8. Following K06, we used a scale radius $\beta$ that
is 1.3 times the dispersion of the best-fit Gaussian: this choice
favours a more accurate description of the outer wings of the images,
and helps to suppress any sub-pixel scale freedom the shapelet fit
might otherwise have. At this point it is assumed that the correct
background level is known; this assumption is addressed in
\S\ref{sec:discussion}.

The GaaP flux is then calculated using eq.~\ref{eq:straightfq}.  and
compared with the exact GaaP flux, as calculated directly from
eq.~\ref{eq:fq2} and plotted in Fig.~\ref{fig:gaap-sersic}. We
repeated this procedure for a range of aperture radii $q$ for each
model, and ran a large set of models of different Sersic index for the
galaxies (0.5 to 4), Moffat index for the PSF (2, 3 and 9), galaxy
effective radius (from completely unresolved to 6 pixels), PSF FWHM (3
to 6 pixels), and location of the galaxy centre with respect to the
pixel grid (center and corner of a pixel, middle of an edge).

The results of this first test are plotted in Fig.~\ref{fig:firstrun},
which show that, provided the Gaussian aperture radius $q$ is within a
factor of two of the shapelet scale of the observed source, the
procedure yields GaaP fluxes that are accurate to 10\% or
better. Going to shapelet order 12 improves the results somewhat (not
shown), but not spectacularly.

While useful, these residuals are still troublingly large.  We now
describe two simple refinements which take account of the residuals to
the shapelet fits to observed source and PSF.

\subsection{Residuals to the observed source expansion}

\begin{figure}
\epsfxsize=\hsize\epsfbox[20 150 575 575]{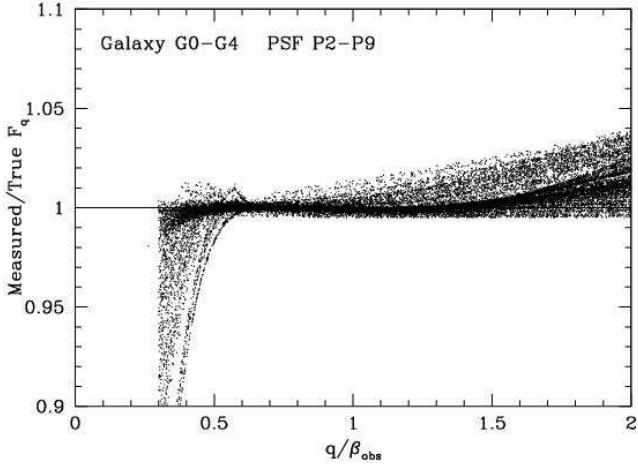}
\caption{As fig.~\ref{fig:firstrun}, but now including the additive correction
  for residual pixel flux to the observed sources' shapelet expansions.}
\label{fig:secondrun}
\end{figure}

The first improvement is to take into account the residual galaxy flux
$R_{ij}$ at pixel $(i,j)$ to the shapelet expansion:
\begin{equation}
R_{ij}=S^\mathrm{obs}(x_i,y_i)-\sum_{a,b}s_{ab}B^{ab}(x_i,y_j).
\end{equation}
Thus far we have simply ignored this flux, but a better approximation
is to include its Gaussian aperture flux (albeit with only an
approximate PSF correction) in the total. Approximating the PSF by a
Gaussian of dispersion $g_\mathrm{psf}$, we can estimate the
residual GaaP flux by convolving $R_{ij}$ with a Gaussian of
dispersion $(q^2-g_\mathrm{psf}^2)^{1/2}$, and computing the
Gaussian-weighted aperture flux of this convolution:
\begin{equation}
\int d{\bf r} \int d{\bf r'} R({\bf r'}) 
{e^{-({\bf r}-{\bf r'})^2/(2q^2-2g_\mathrm{psf}^2)} \over 
2\pi (q^2-g_\mathrm{psf}^2)} e^{-r^2/2q^2}
\end{equation}
which, after changing the order of integration can be written (as a
generalization of eq.~\ref{eq:fq2}) as the
aperture integral
\begin{equation}
\int d{\bf r} R({\bf r}) {q^2\over2q^2-g_\mathrm{psf}^2} e^{-r^2/(4q^2-2g_\mathrm{psf}^2)}.
\end{equation}
Thus we add
\begin{equation}
F_\mathrm{res} = 
\sum_{i,j}R_{ij} {q^2\over2q^2-g_\mathrm{psf}^2}
e^{-(x_i^2+y_j^2)/(4q^2-2g_\mathrm{psf}^2)}
\label{eq:fqres}
\end{equation}
to the result obtained from eq.~\ref{eq:straightfq}.  For the
simulated Sersic profiles this term can be of order 10\% of $F_q$,
particularly for very peaked images. The effect of incorporating this
correction can be seen in Fig.~\ref{fig:secondrun}.

\subsection{Residuals to the PSF model}

\begin{figure}
\epsfxsize=\hsize\epsfbox[20 150 575 575]{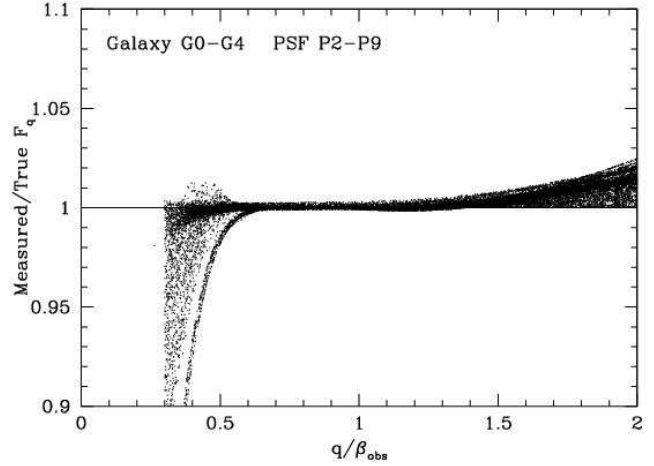}
\caption{As fig.~\ref{fig:secondrun}, but now furthermore including the
  multiplicative correction
  for residual pixel flux to the PSFs' shapelet expansions.}
\label{fig:thirdrun}
\end{figure}

The second effect is caused by residuals to the PSF's shapelet
expansion, which can introduce a multiplicative scaling error during
the deconvolution. (Essentially, if some of the flux of the PSF is
missed in the shapelet model then deconvolution by this underestimated
PSF will result in an overestimated galaxy flux.) To estimate the
effect of this missed PSF flux, we compute the correction factor for
a Gaussian galaxy.

For a Gaussian galaxy, of dispersion $g$ and unit total flux, the GaaP flux
$F^q$ is equal to (see eq.~\ref{eq:fq2})
\begin{equation}
\int d{\bf r}\,
\frac12 {e^{-r^2/2g^2}\over2\pi g^2} e^{-r^2/4q^2} = {q^2\over 2q^2+g^2}.
\label{eq:fqgaus}
\end{equation}
However, if a part $P_\mathrm{res}(x,y)$ of the PSF $P$ was not included
in the shapelet model, then to first order in these residuals $F_q$
will have been overestimated by an amount
\begin{equation}
\int d{\bf r}\, \left[P_\mathrm{res} \otimes^{-1} P \otimes G(q) \otimes
G(g)\right] e^{-r^2/2q^2} .
\label{eq:decon1}
\end{equation}
Here $G(\sigma)$ is a Gaussian of
dispersion $\sigma$, $\otimes$ denotes convolution and $\otimes^{-1}$
deconvolution. We can estimate this integral by
replacing the PSF $P$ by a
Gaussian whose dispersion $g_\mathrm{psf}$ is chosen to be a
reasonable match to the PSF size. Combining all Gaussian
(de)convolutions reduces eq.~\ref{eq:decon1} to
\begin{equation}
\int d{\bf r}\, \left[P_\mathrm{res} \otimes 
G\left(\sqrt{q^2-g_\mathrm{psf}^2+g^2}\right)\right] e^{-r^2/2q^2}
\label{eq:decon2}
\end{equation}
or, after the by now customary swapping of the order of integration,
\begin{equation}
{q^2\over 2q^2+g^2-g_\mathrm{psf}^2} 
\int d{\bf r}\, P_\mathrm{res} e^{-r^2/(4q^2+2g^2-2g_\mathrm{psf}^2)}  .
\label{eq:decon3}
\end{equation}
The ratio of eqs.~\ref{eq:decon3} and \ref{eq:fqgaus} is
then an estimate of the fractional error that results from the
incomplete PSF model. Correcting for this error is accomplished by
dividing $F_q+F_\mathrm{res}$ (from eqs.~\ref{eq:straightfq} and
\ref{eq:fqres}) by
\begin{equation}
f_\mathrm{psf}=1+{2q^2+g^2\over 2q^2+g^2-g_\mathrm{psf}^2}
\int d{\bf r}\, P_\mathrm{res} e^{-r^2/(4q^2+2g^2-2g_\mathrm{psf}^2)} .
\end{equation}
In practice we estimate $g$, the intrinsic Gaussian size of the source, 
from the best-fit Gaussian radius of the observed source $g_\mathrm{obs}$ as 
$g^2\simeq g_\mathrm{obs}^2-g_\mathrm{psf}^2$. 

In figure~\ref{fig:thirdrun} we show the result of implementing these
two corrections: the GaaP flux is now systematically correct to better
than 2\%, over a wide range of Sersic index and PSF size, and for GaaP
aperture radii between half and twice the shapelet scale of the
observed images. The error is below 1\% for $0.6<q/\beta<1.6$, with
the worst errors arising for PSFs and galaxies that are only a few
pixels across.

\begin{figure}
\epsfxsize=\hsize\epsfbox[20 150 575 575]{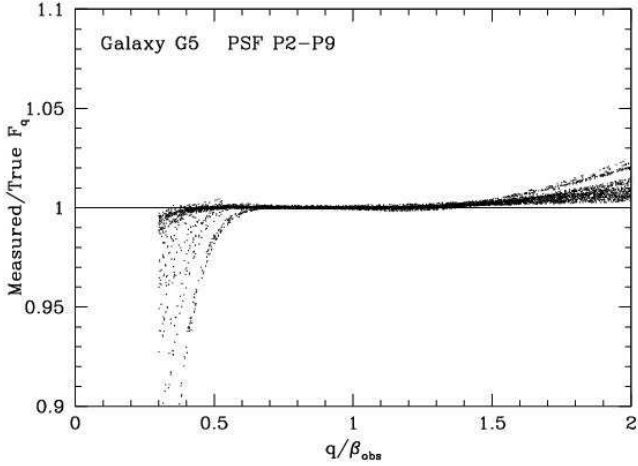}
\caption{As Fig.~\ref{fig:thirdrun}, but for a two-armed spiral
  galaxy}
\label{fig:fqallcor-spiral}
\end{figure}

\begin{figure}
\epsfxsize=\hsize\epsfbox[20 150 575 575]{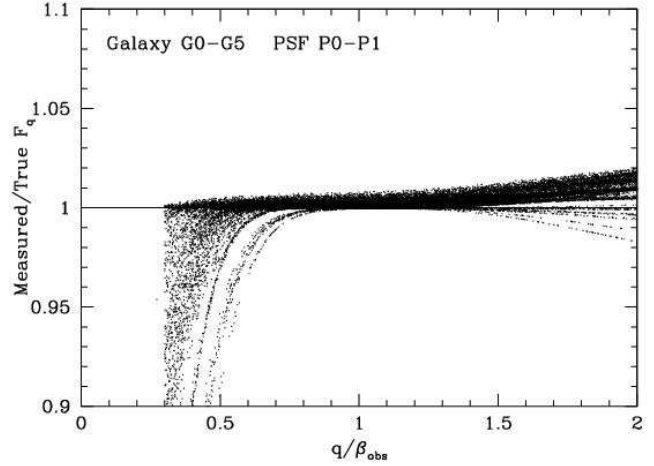}
\caption{As Fig.~\ref{fig:thirdrun}, but for PSFs with diffraction
  rings. PSFs 0 and 1 are the same except that PSF 1 includes four
  diffraction spikes which together contain 10\% of the flux. }
\label{fig:fqallcor-airy}
\end{figure}

\begin{figure}
\epsfxsize=\hsize\epsfbox[20 150 575 575]{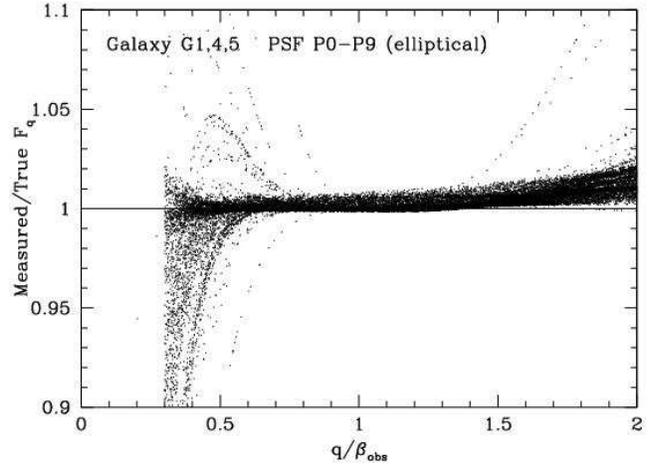}
\caption{As Fig.~\ref{fig:thirdrun}, but for elliptical PSFs of axis
  ratio 3:2. PSF and galaxy types have the same meaning as before.}
\label{fig:fqallcor-ell}
\end{figure}

\begin{figure}
\epsfxsize=0.49\hsize\epsfbox{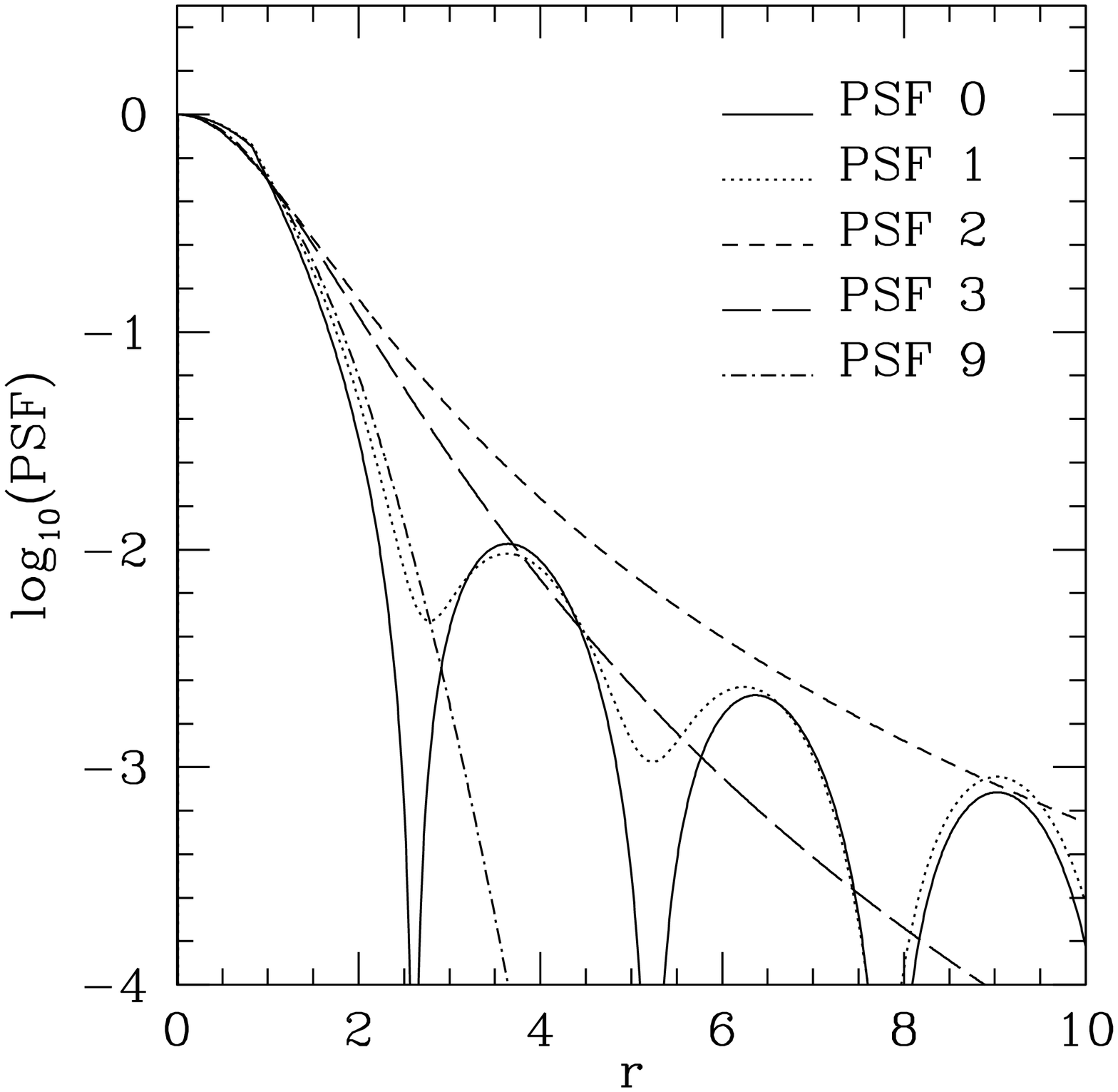}
\epsfxsize=0.49\hsize\epsfbox{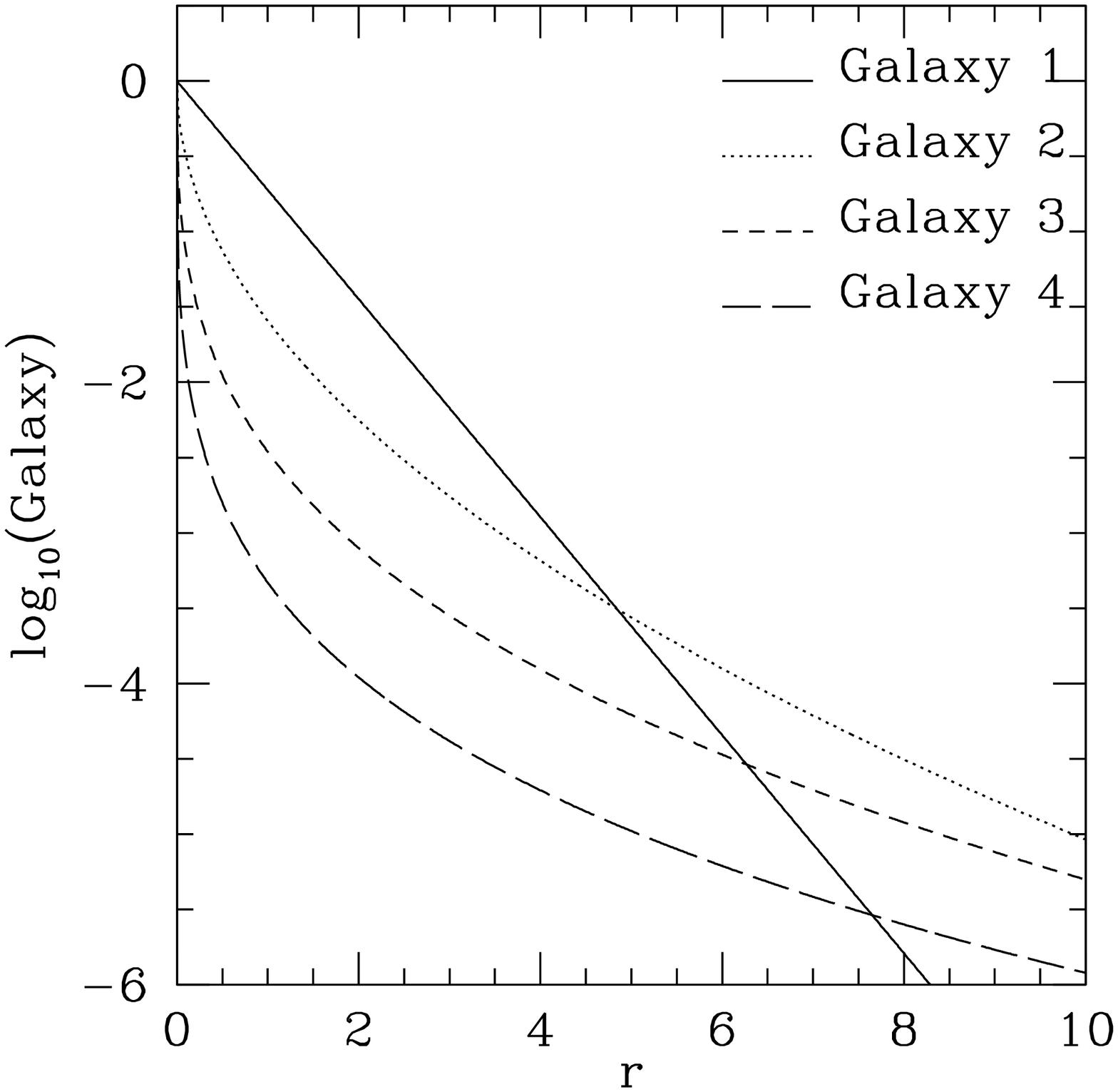}
\caption{Azimuthally averaged profiles of the PSF (left) and galaxy
(right) models used in the simulations. 
The PSF models are normalized to have FWHM=2, the galaxy
models have effective (i.e., half-light) radius equal to 1.}
\label{fig:models}
\end{figure}

\begin{figure}
\epsfxsize=\hsize\epsfbox[74 195 447 693]{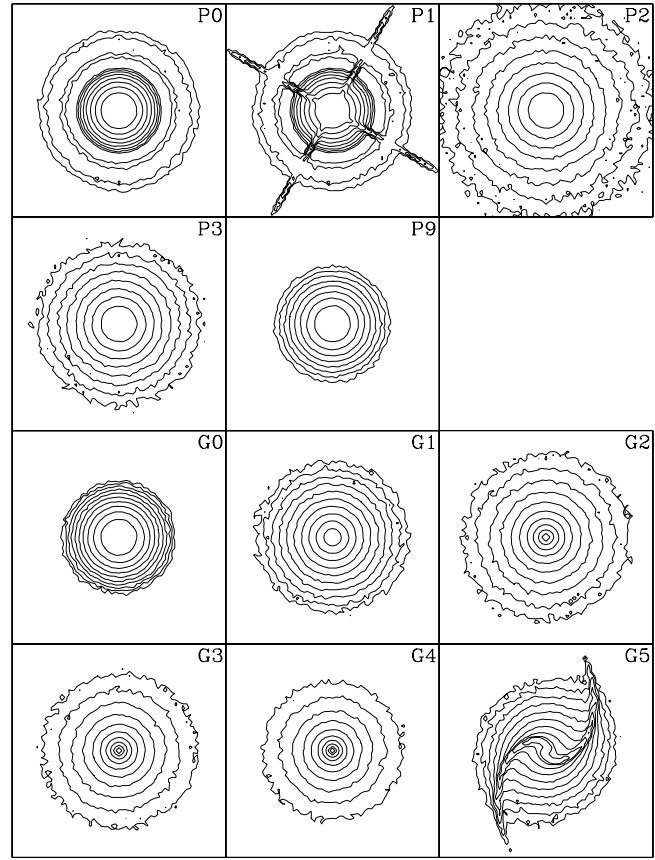}
\caption{Contour plots of the models used in the simulations. Top two
rows: PSF models 0--9; bottom two rows: Galaxy models 0--5. Contours
levels differ by a factor of two. The box sizes are six times the FWHM
for the PSFs and twelve times the effective radius for the galaxies.
See Table~\ref{tab:types} for detailed descriptions of these models.}
\label{fig:modcont}
\end{figure}

\subsection{More complex morphologies}

Having established that the method works for circular, regular
galaxies, we extended the tests to more complicated shapes. In
Fig.~\ref{fig:fqallcor-spiral} we show the $F_q$ measurements for a
simulated logarithmic spiral galaxy, for the same PSFs as above; the
model used is obtained by replacing the random azimuths $\theta$ of
stars in a circular exponential disk model by
\begin{equation}
\theta\rightarrow 
\theta-\frac12\cos\left(2(\theta-\ln r)\right).
\label{eq:spiral}
\end{equation}
(The amplitude
$\frac12$ is chosen to give maximally sharp spiral arms.)  

In Fig.~\ref{fig:fqallcor-airy} we show the results for all galaxy
models and with two PSFs that mimick diffraction-limited PSFs. PSF
type 0 is defined as
\begin{equation}
\mathrm{PSF0}(r)=\left\{
\begin{array}{ll}
{\sin ^2r/ r^2}&\qquad r<1;\\
{\sin ^2r/ r^3}&\qquad r>1.
\end{array}
\right.
\label{eq:pseudoairy}
\end{equation}
and PSF type 1 is the same except that 10\% of the flux is put in four
thin linear 'diffraction spikes' along which the density varies as
$\mathrm{PSF0}(r/2^{1/2})$. 
Even here
the GaaP flux is recovered to better than 1\% for aperture $q$ between
0.7 and 1.5 times $\beta_\mathrm{obs}$. In the case where the PSF FWHM
is 2 pixels (not shown) the differences can rise to 5\%.

Finally, a series of simulations were analysed for circular galaxies
that were convolved with elliptical PSFs (axis ratio 3:2). Also here,
the results are satisfactory (see Fig.~\ref{fig:fqallcor-ell}).

We therefore conclude that the method works.

More details of the relation between the residuals and the model
parameters are given in colour figures included with the on-line
material.  We close this section with plots (Fig.~\ref{fig:models} and
~\ref{fig:modcont}) of the PSF and galaxy models used in the
simulations, and remark that even though the shapelet formalism is
built around a Gaussian `parent' flux distribution, it can be used
succesfully to do photometry on much more extended
sources. Nevertheless, it may be profitable to consider expansions
around more galaxy-like parent functions in the future.

\begin{figure}
\epsfxsize=\hsize
\epsfbox{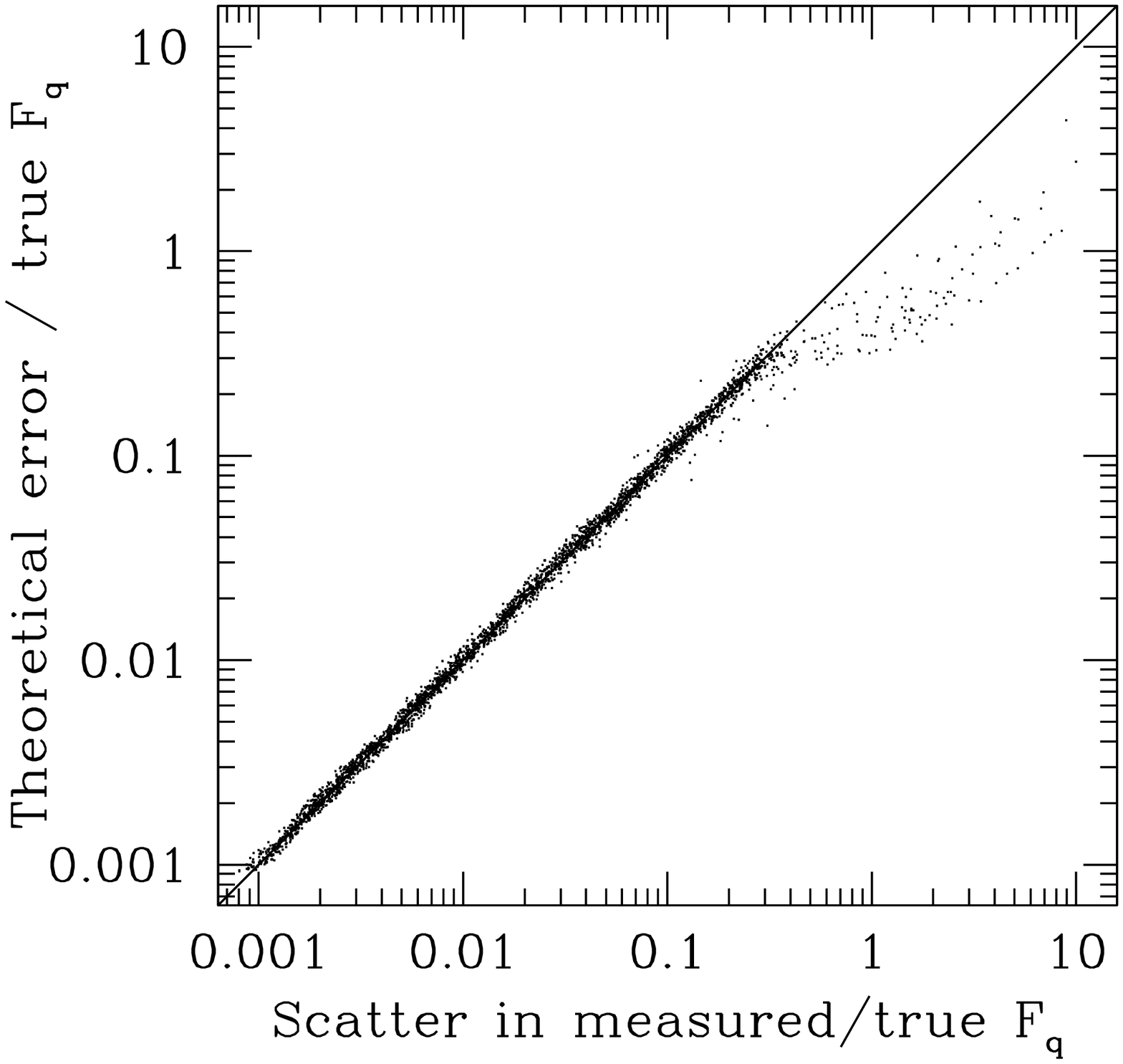}
\epsfxsize=\hsize
\epsfbox{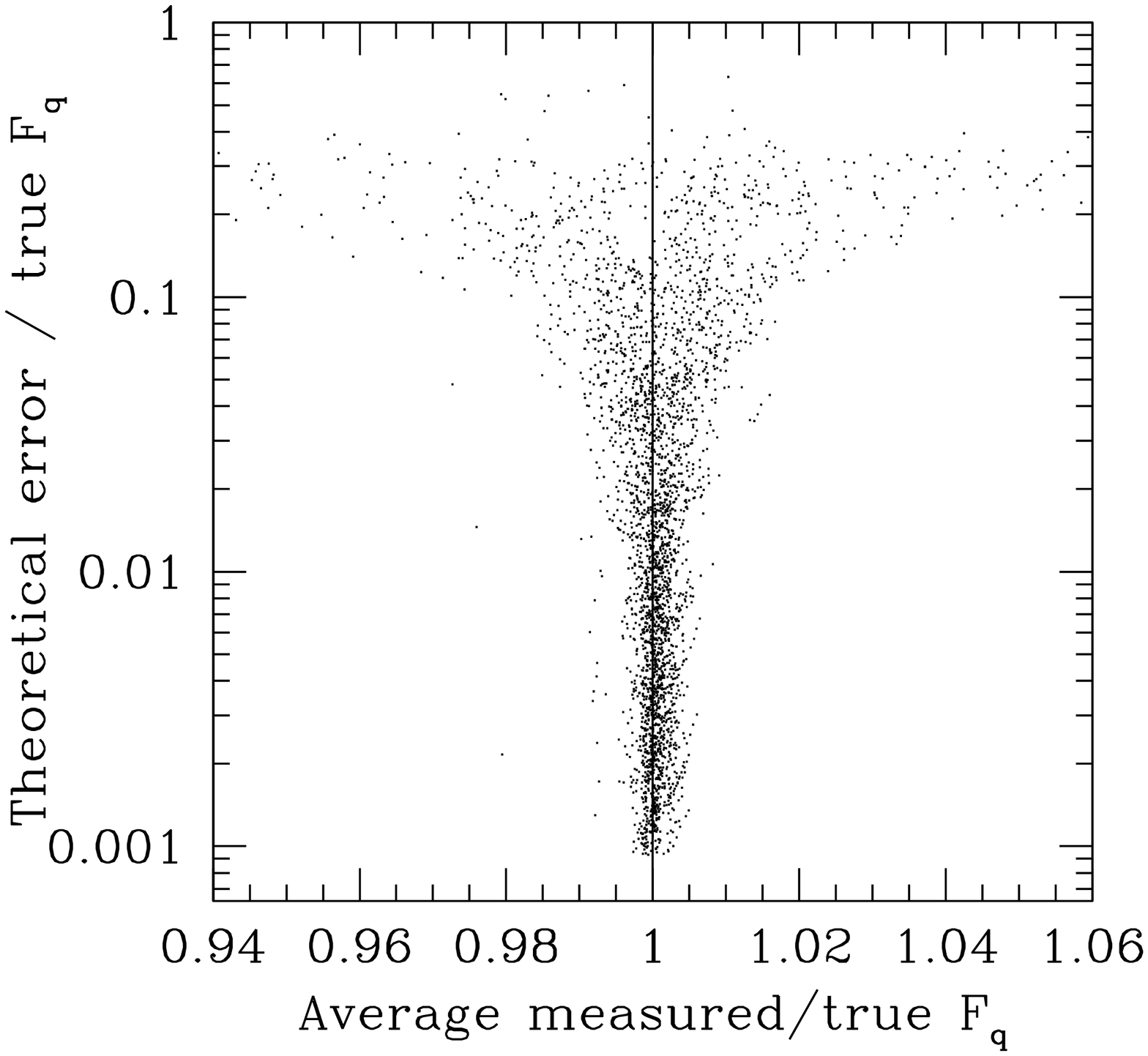}
\caption{Top: The scatter in measured GaaP fluxes from simulated images,
  $\sqrt{\hbox{Var}(F_q)}/F_q$, 
  versus the theoretical prediction of eq.~\ref{eq:varfq}. Each point
  represents a set of 100 noisy realizations of a particular
  combination of PSF and Galaxy models from Table~\ref{tab:types}, with
  aperture radius $q$ between 0.6 and 1.4 times $\beta_\mathrm{obs}$.
  The scatter is adequately predicted for signal-to-noise ratios above
  about three.
  Bottom: the average GaaP flux for each set of 100 realizations,
  versus the rms scatter, showing that there is no noise-related bias
  in the results.
}
\label{fig:noise}
\end{figure}

\section{Noise}
\label{sec:noise}

A similar set of tests can be used to investigate the effect of noise
on the GaaP fluxes: we simply add a constant Gaussian noise level to
all pixels in simulated galaxy images, as appropriate for
background-limited images, and repeat the analysis many times. In
Fig.~\ref{fig:noise} we show the comparison between the errors 
prediction from eq.~\ref{eq:varfq}, and the scatter from a large
number of realizations, using the models of Figs.~\ref{fig:firstrun}
to \ref{fig:fqallcor-airy}. 

For adequately resolved (FWHM 3 pixels or greater), or more
Moffat-like, PSFs, the theoretical error bar adequately predicts the
scatter in the measurements, and any bias in the results is neglegibly
small.

\begin{figure}
\epsfxsize=\hsize\epsfbox{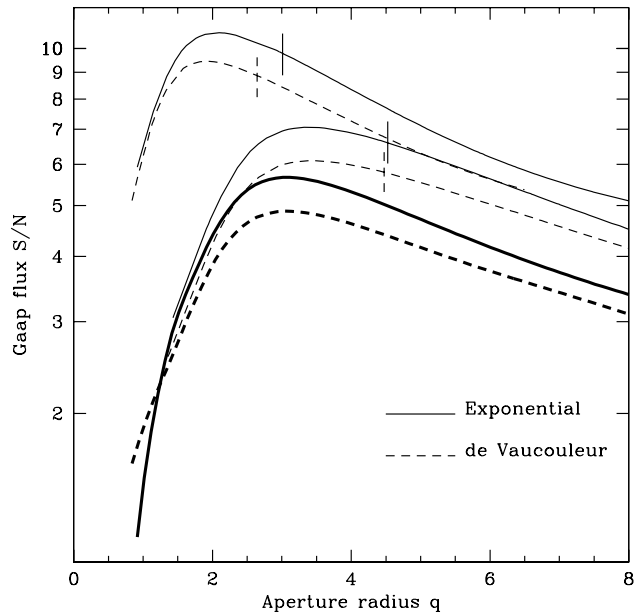}
\caption{The effect of aperture radius on S/N of the GaaP fluxes and
  colours. The thin solid curves show the S/N of the GaaP flux of an
  exponential disk, $R_e=2$ pixels, convolved with Moffat $\beta_m=3$
  PSFs of FWHM 3 and 6 pixels (top and middle curves---tick marks
  indicate the shapelet scale radius used to model the sources). The
  heavy line shows the S/N of the ratio of the fluxes. For this
  calculation is was assumed that both images have the same pixel
  noise of 1\% of the total flux of the galaxy. The dashed line shows
  the same result for a de Vaucouleur model galaxy.}
\label{fig:sncolour}
\end{figure}

Deconvolution will always amplify photon noise, so it is important to
try to choose an optimal aperture radius. This is addressed in
Fig.~\ref{fig:sncolour}, which shows how the S/N of the Gaap fluxes
and of their ratio varies with aperture radius. In this calculation we
assume that a galaxy is observed with PSFs of two different sizes, so
that a given choice of aperture radius $q$ requires a different degree
of (de)convolution. The ratio of the two GaaP fluxes at a given $q$
represents a colour measurement, with the S/N ratios on the fluxes
adding in inverse quadrature. As the figure shows, there is an optimal
aperture to use, but the maximum is quite broad, so that useful
colours can be measured for a wide range of $q$. It certainly is not
optimal to smooth the best-seeing image to the worst one.

\begin{figure}
\epsfxsize=\hsize\epsfbox[20 150 575 575]{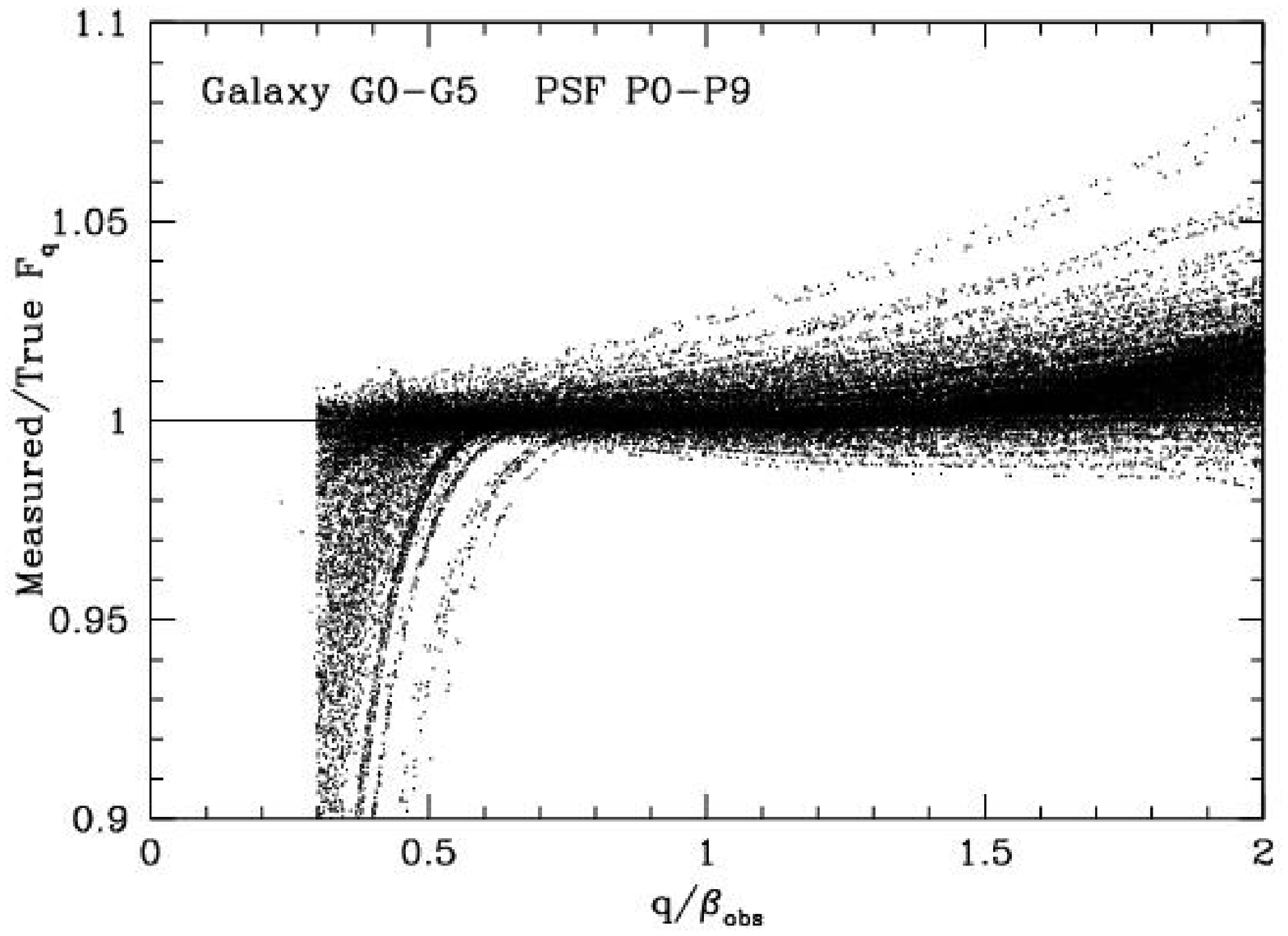}
\caption{The same simulation as in Fig.~\ref{fig:thirdrun}, except
  that before the shapelet fitting a locally-fitted background level,
  determined from an annulus between radius $8\beta$ and $11\beta$ is
  subtracted. Particularly the 'wingy' PSFs 0, 1 and 2 are affected by
  this background error.}
\label{fig:bgsub}
\end{figure}

\section{Discussion}
\label{sec:discussion}

We have shown that the proposed technique is an effective way to
generate well-defined aperture fluxes, with accurate error estimates,
that are independent of PSF and pixel scale of the observations.
In all cases with PSFs that are well-resolved (FWHM at least 3 pixels)
systematic errors are below a percent.

The advantage of this approach is that it allows multi-colour
photometry to be done on a distributed dataset. All that is needed is
a set of object coordinates and aperture radii, and PSF models for
each image: then tables of GaaP fluxes can be extracted from each
separate image and simply merged into a multi-colour catalogue. The
ratios of the GaaP fluxes of aperture radius $q$ are then true
integrated colours, corresponding to the ratio of the
Gaussian-weighted fluxes of the intrinsic sources with the same weight
function $\exp(-r^2/4q^2)$. It should be stressed that the GaaP fluxes
are not total fluxes.

This method does not absolve one from knowing the correct background
level and zeropoint of each image. This can be difficult with extended
galaxies and PSFs, and any flux that is missed in either will have
repercussions for the measured aperture fluxes. Provided the images
are well flat-fielded, it is possible to obtain decent local
background estimates in annuli around each source or star. This will
include some source flux from the wings of the flux profile, but for
sufficiently large annuli the effect is small. We have tested the
effect with the simulations described in \S\ref{sec:tests}, by
subtracting a background level (determined as the median flux in the
pixels between $8\beta$ and $11\beta$ from the centre of the source)
before performing the shapelet decomposition. The resulting GaaP
fluxes, including all corrections for pixel residuals, are shown in
Fig.~\ref{fig:bgsub}. Not surprisingly, larger apertures are more
affected by background level errors, particularly in the PSF. Overall,
while the effect is visible, it is not catastrophic: moreover to first
approximation the same amount of 'wing' will be subtracted in
different bands so that the overall colours should be affected even
less.

We have argued that our procedure is advantageous in the context of
large multi-band imaging surveys.  The alternative approach of
convolving entire images to a common PSF, followed by matched-aperture
photometry, is less practical in that it requires much more
information exchange between the different images (it may also be more
computer-intensive as it requires all pixels to be convolved). In the
context of huge multi-band surveys, in which different bands may be
observed with different instruments or processed in different places,
the logistics of these exchanges are complex.

A different, catalogue-based solution to the problem of deriving
accurate colours would be to generate model total fluxes for all
sources, e.g. by fitting detected sources to combinations of
PSF-convolved Sersic profiles, as done for the Sloan Digital Sky
Survey (e.g., Adelman-McCarthy et al.~\cite{sdssdr4}). This can also
be done image by image, and also yields fluxes that are
PSF-independent, but the results will only be valid if the model is
appropriate. Few-parameter models are a priori unlikely to yield
percent-level accurate total fluxes, and any colour gradients in the
source will give rise to wavelength-dependent systematic errors to
such a simple fit. By contrast, the method proposed here makes
virtually no assumptions on the galaxy shapes.

The technique described in this paper can be used to compare any set
of images of a piece of sky: while we have concentrated on colour
measurements, another application would be to compare images taken at
different epochs to measure time variability. Most time-varying
sources are point-like, in which case the more straightforward
approach of comparing PSF-fitting fluxes might be more appropriate,
but a possible application would be variability of embedded point
sources such as galactic nuclei or distant supernovae.

\section{Summary}
\label{sec:summary}

We have defined a PSF-independent aperture flux, the GaaP
(Gaussian-aperture and PSF) flux, and showed how it can be extracted
from astronomical images provided the point spread function is
known. An extensive series of tests on simulated images demonstrates
that the level of systematic residual is at or below the 1\% level,
for aperture sizes that are up to a factor of two larger or smaller
than the observed source. Implementation of these ideas into robust
data processing pipelines will allow accurate colours to be extracted
from large wide-field imaging surveys, virtually free from residual
PSF and pixel scale effects.

\begin{acknowledgements}
  This article was completed during a stay at the Kavli Institute for
  Theoretical Physics in Santa Barbara, and their hospitality and
  support is gratefully acknowledged. This research was supported in
  part by the National Science Foundation under grant no.~PHY99-07949,
  the Netherlands Organization for Scientific Research (NWO) and the
  Leidsch Kerkhoven-Bosscha Fonds.  I thank U.~Hopp, R.~Saglia and
  G.~Verdoes for helpful comments on an earlier version of the
  manuscript.
\end{acknowledgements}

\Online
\appendix

\section{Colour-coded residual plots}

This Appendix shows the relation between the residuals in $F_q$
measurements and the model parameters. The data plotted are identical
to those of Figs.~\ref{fig:firstrun} to~\ref{fig:fqallcor-ell}, but
the points have been colour-coded according to different model
parameters. This helps to identify the most discrepent cases.

As a key to the figures, a summary of all PSF and galaxy models tested
is given in Table~\ref{tab:types}.

\begin{figure*}
\epsfxsize=\hsize\epsfbox{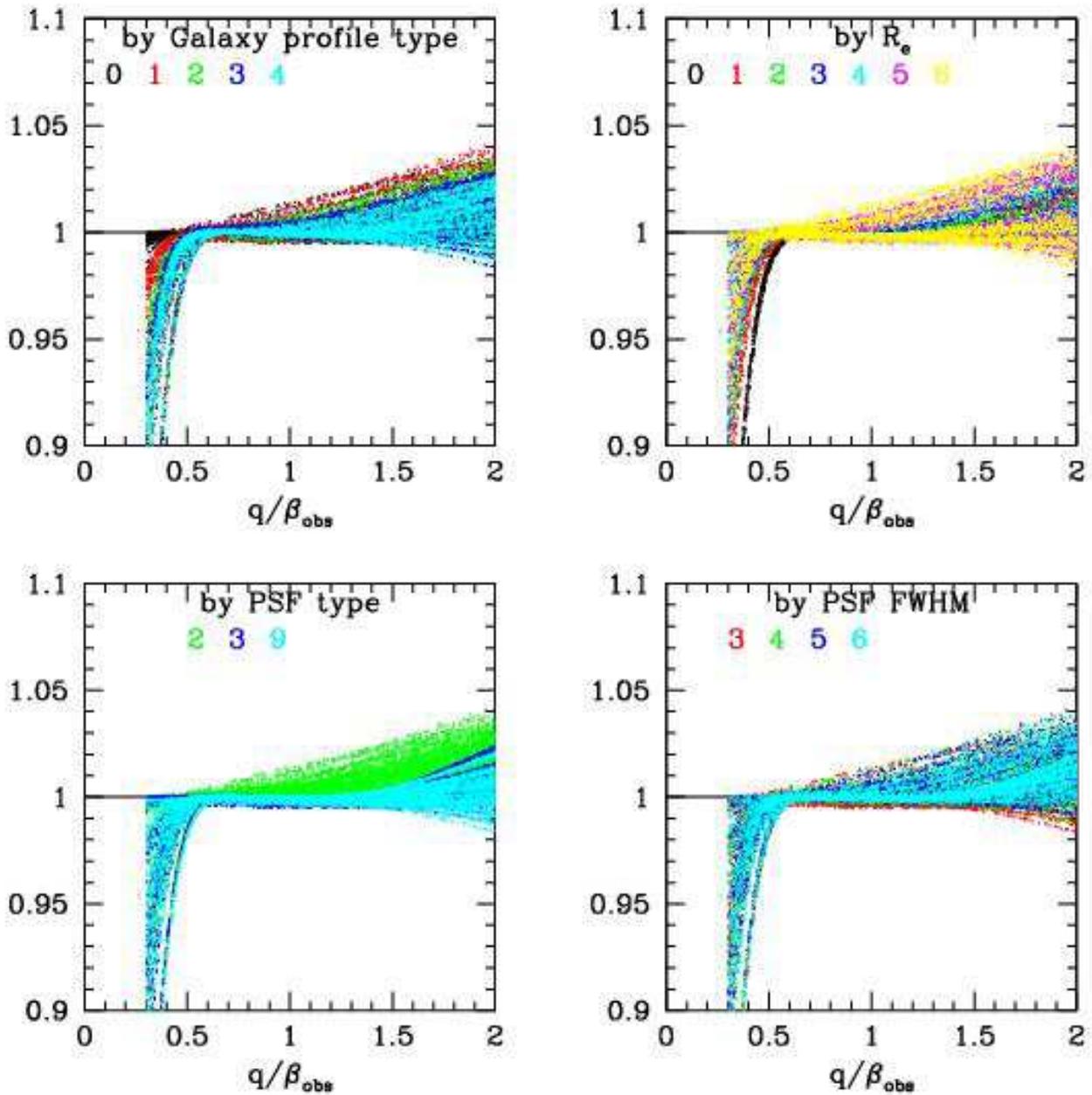} 
\caption{The results of the simulations plotted in
Fig.~\ref{fig:firstrun}, shown here as four panels. Each panel shows
the same results, but colour-coded according to a different input
parameter: top left, Galaxy type; top right, Galaxy effective radius;
bottom left, PSF type; bottom right, PSF FWHM. The colour used to plot
the results for each parameter value is shown in each panel, and the
model types are identified in Table~\ref{tab:types}.}
\label{fig:colourpanel}
\end{figure*}

\begin{figure*}
\epsfxsize=\hsize\epsfbox{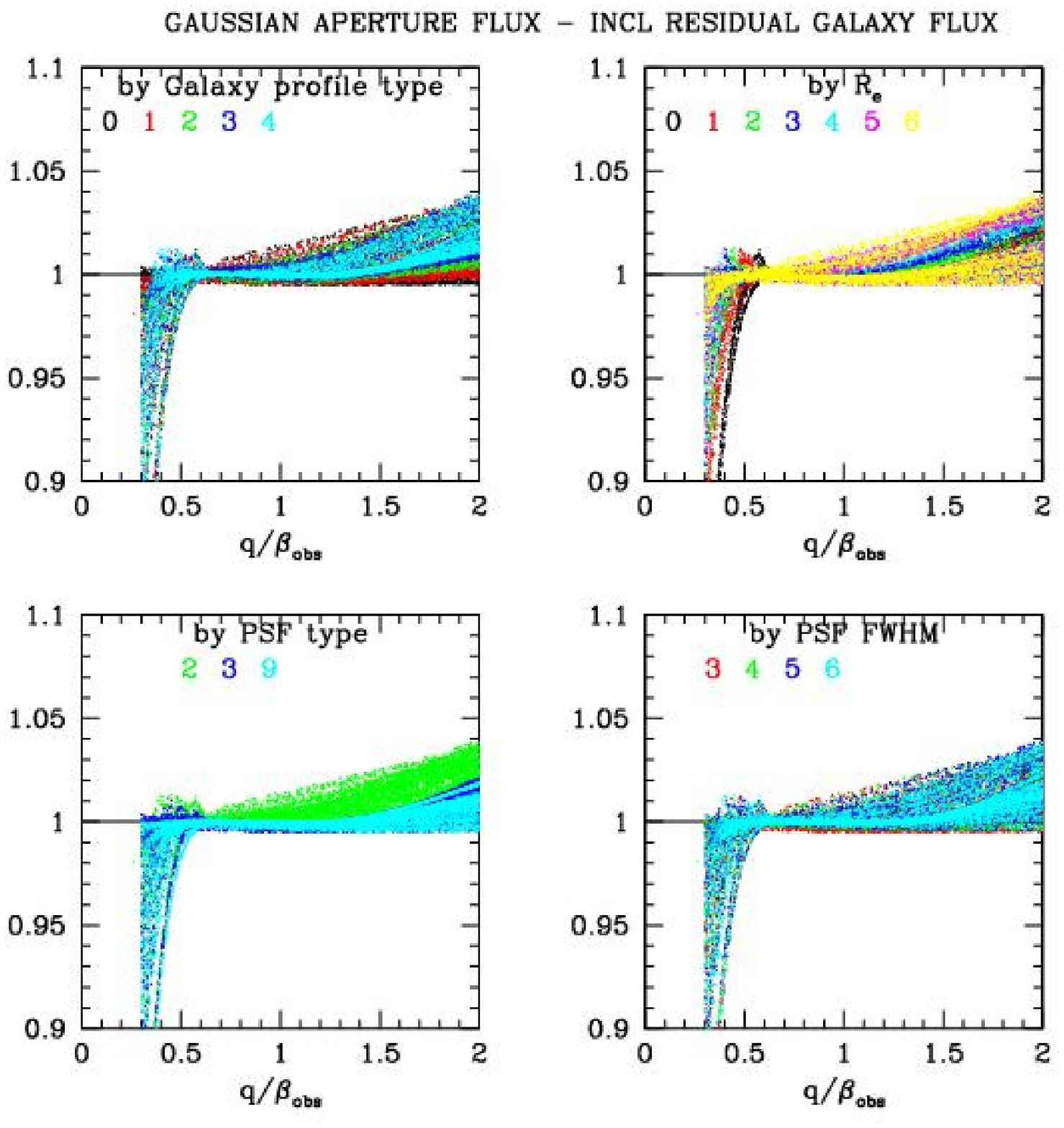} 
\caption{As Fig.~\ref{fig:colourpanel}, but for the results of the
simulations plotted in Fig.~\ref{fig:secondrun}.}
\end{figure*}

\begin{figure*}
\epsfxsize=\hsize\epsfbox{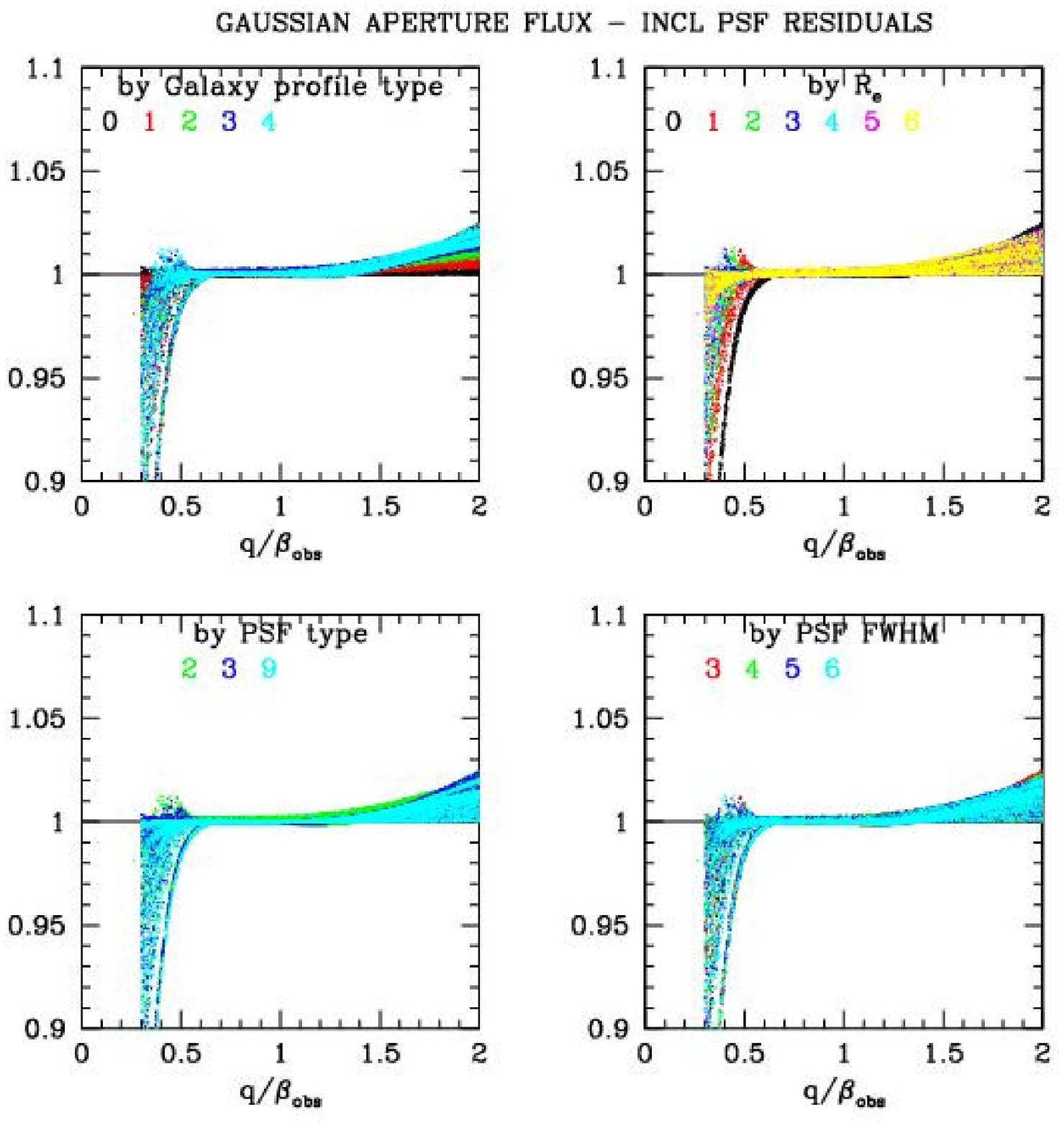} 
\caption{As Fig.~\ref{fig:colourpanel}, but for the results of the
simulations plotted in Fig.~\ref{fig:thirdrun}.}
\end{figure*}

\begin{figure*}
\epsfxsize=\hsize\epsfbox{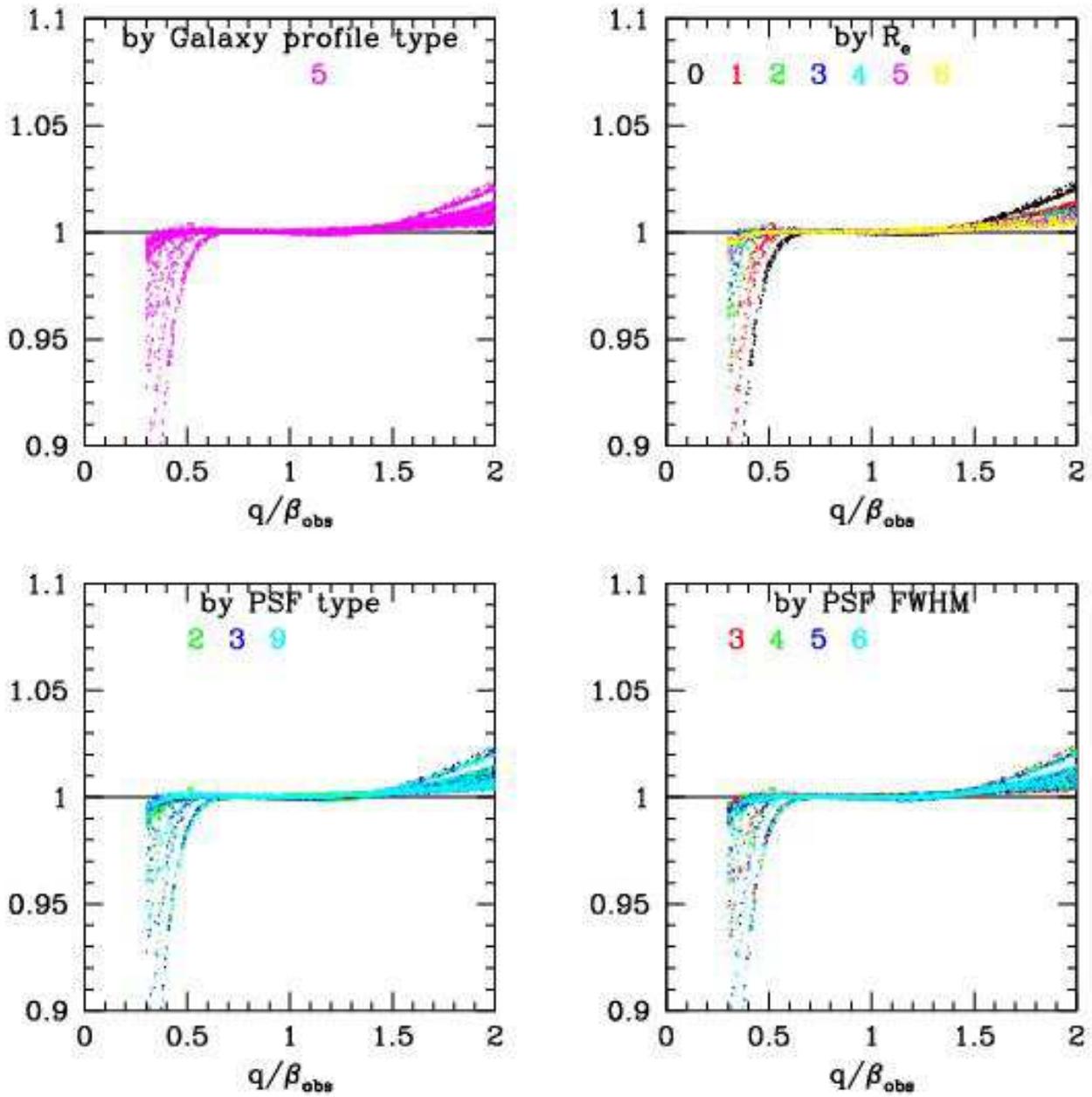} 
\caption{As Fig.~\ref{fig:colourpanel}, but for the results of the
simulations plotted in Fig.~\ref{fig:fqallcor-spiral} (spiral galaxy model).}
\end{figure*}

\begin{figure*}
\epsfxsize=\hsize\epsfbox{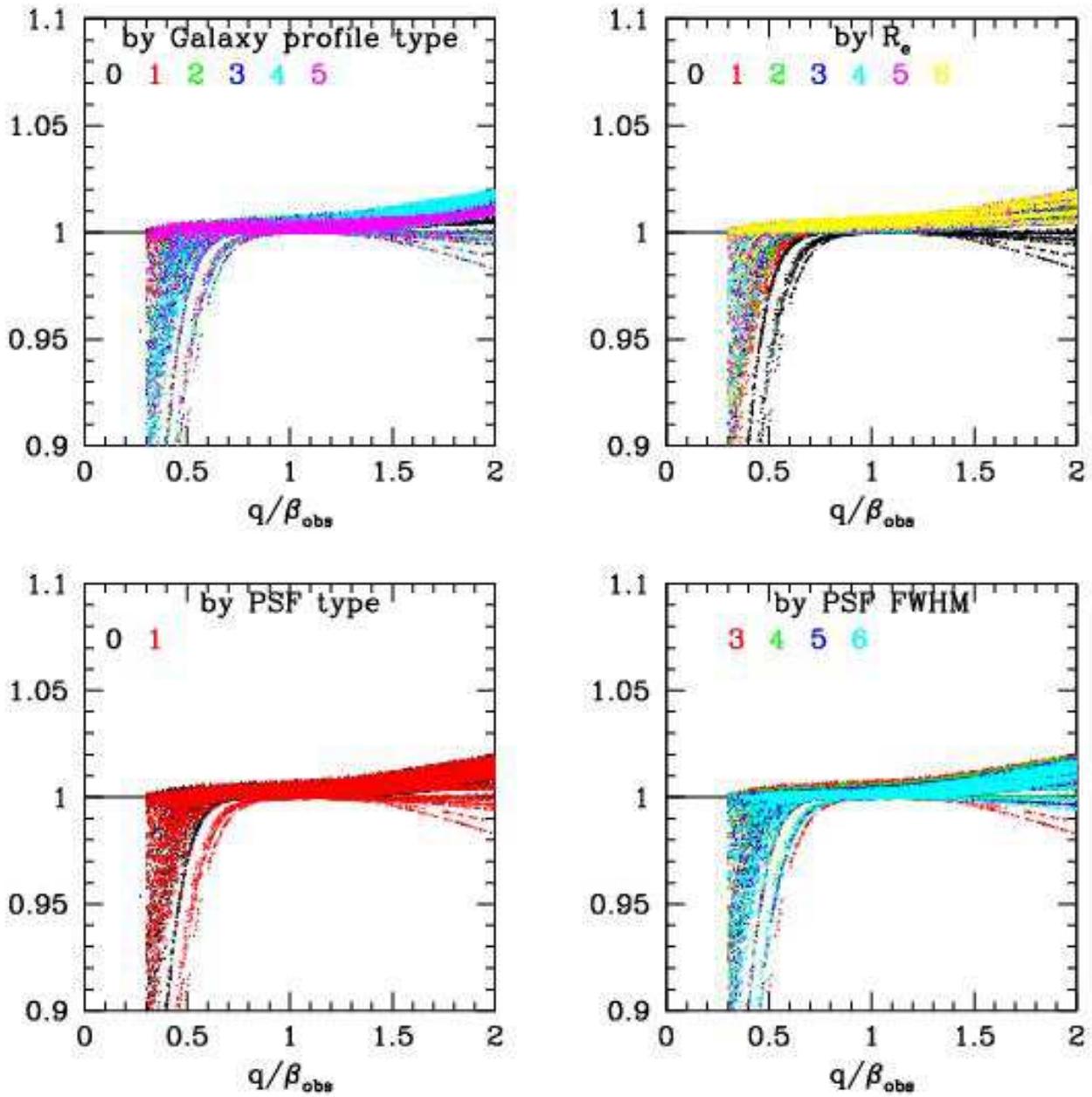} 
\caption{As Fig.~\ref{fig:colourpanel}, but for the results of the
simulations plotted in Fig.~\ref{fig:fqallcor-airy} (Airy
function-like PSFs).}
\end{figure*}

\begin{figure*}
\epsfxsize=\hsize\epsfbox{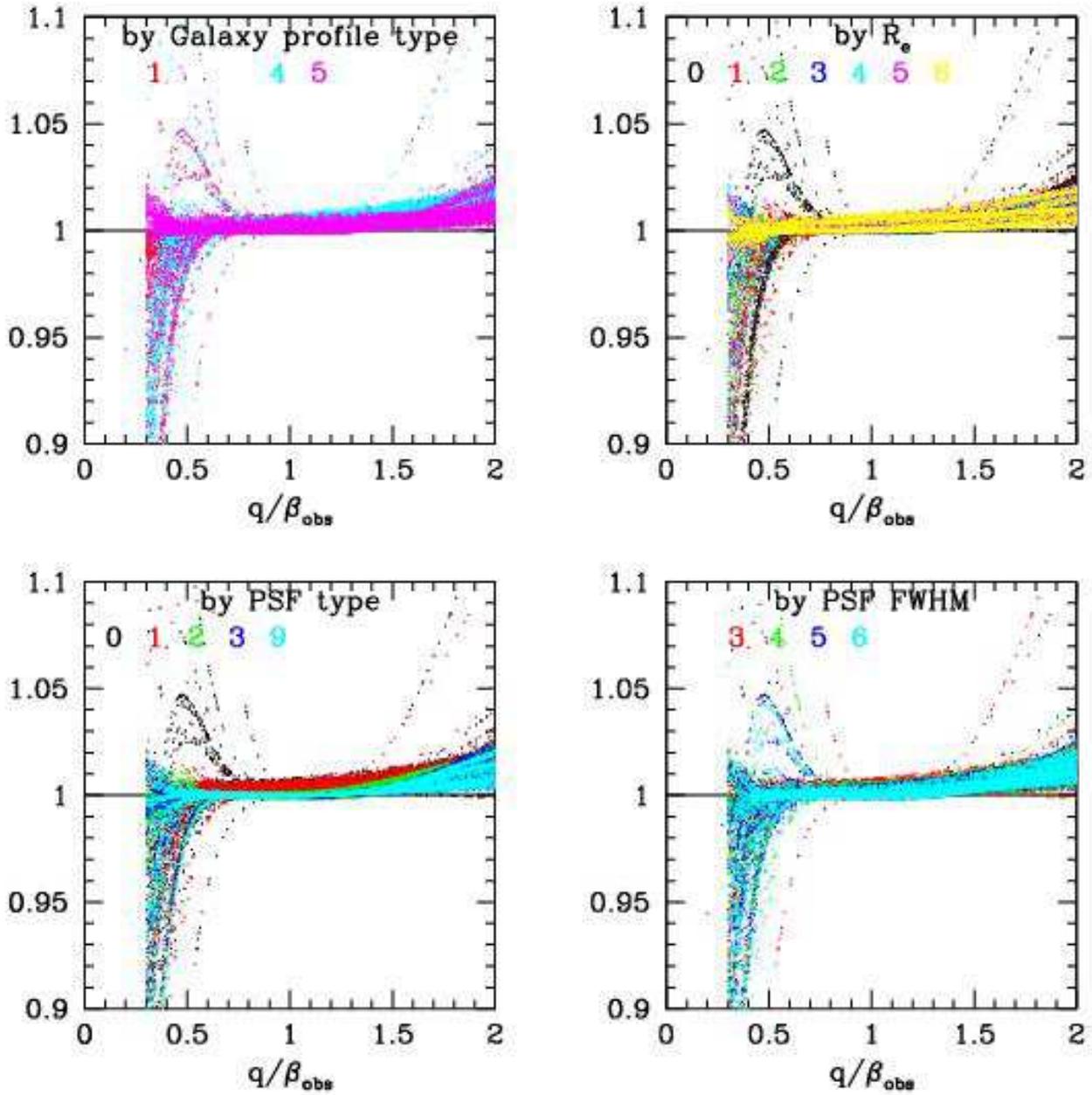} 
\caption{As Fig.~\ref{fig:colourpanel}, but for the results of the
simulations plotted in Fig.~\ref{fig:fqallcor-ell} (elliptical PSFs).}
\end{figure*}

\begin{figure*}
\epsfxsize=\hsize\epsfbox{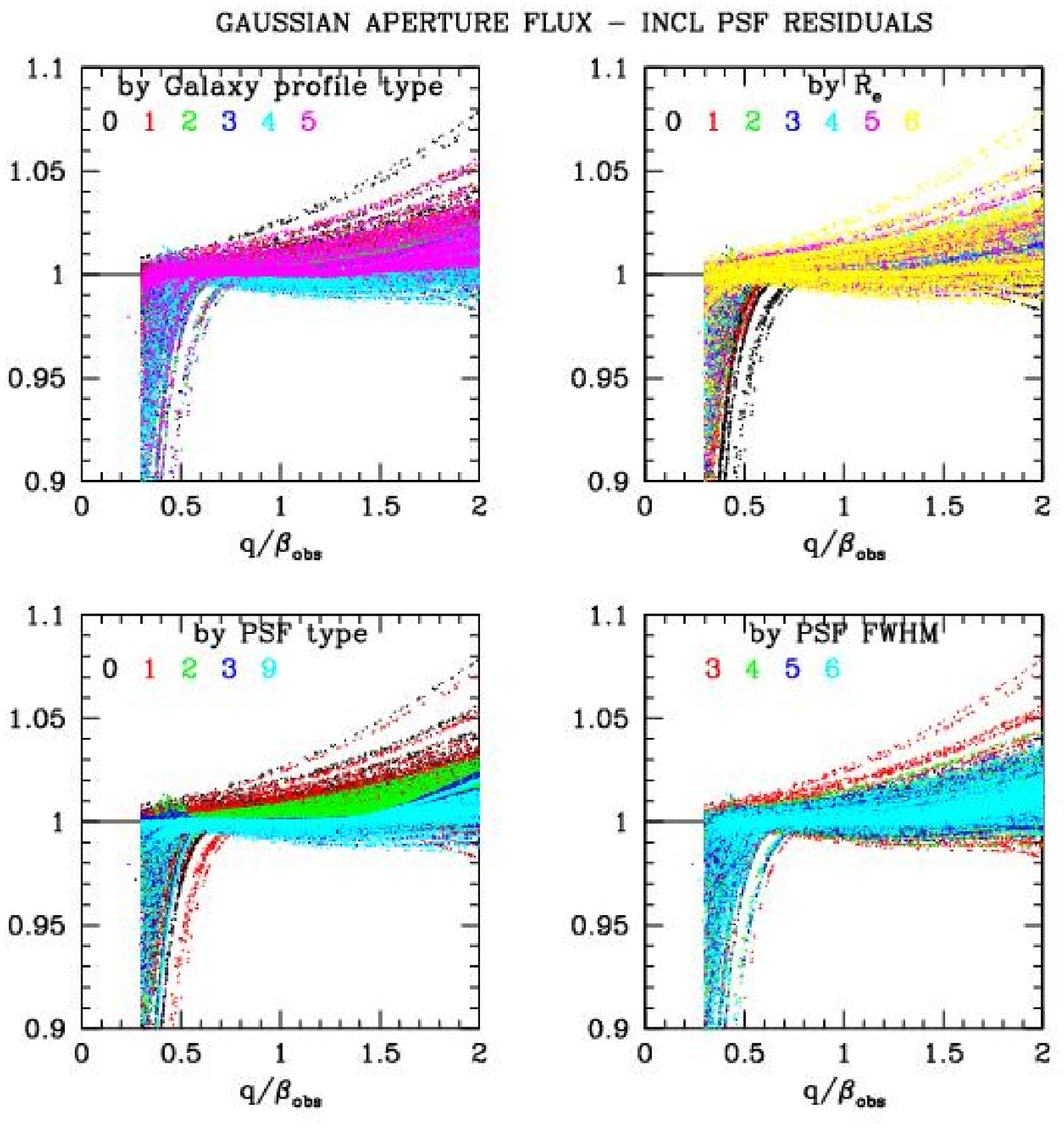}
\caption{As Fig.~\ref{fig:colourpanel}, but for the results of the
  simulations plotted in Fig.~\ref{fig:bgsub} (locally fitted
  background subtracted from source and PSF).}
\end{figure*}


\begin{thebibliography}{}

\bibitem[1972]{a+s} Abramowicz, M., \& Stegun, I. A. 1972, 
Handbook of Mathematical Functions (New York: Dover) First citation in article

\bibitem[2006]{sdssdr4} Adelman-McCarthy, J.K.\ et al.\ 2006, \apjs, 162, 38

\bibitem[1957]{Baum1957} Baum, W.A.\ 1957, \aj, 62, 6

\bibitem[2006]{cfhtls3} Brodwin, M., Lilly, S.J., Porciani, C.,
  McCracken, H.J., Le Fèvre, O., Foucaud, S., Crampton, D., \&
  Mellier, Y.\ 2006, \apjs, 162, 20

\bibitem[2006]{fires2} F\"orster Schreiber, N.M. et al.\ 2006, \aj,
  131, 1891

\bibitem[2003]{fdf} Heidt, J. et al.\ 2003, \aap, 398, 49

\bibitem[2006]{Kuijken2006}Kuijken, K.\ 2006, \aap, in press (K06)

\bibitem[2003]{fires1} Labb\'e et al.\ 2003, \aj, 125, 1107

\bibitem[2003]{Refregier2003}Refregier, A.\ 2003,  \mnras,  338,  35 (R03)

\bibitem[1968]{Sersic1968}Sersic, J.~L., Atlas de galaxias Australes

\end{thebibliography}
\end{document}